\documentclass{article} 
\usepackage{iclr2027_conference,times}

\usepackage{amsmath,amsfonts,bm}

\def\eqref#1{equation~\ref{#1}}

\def\1{\bm{1}}

\DeclareMathAlphabet{\mathsfit}{\encodingdefault}{\sfdefault}{m}{sl}
\SetMathAlphabet{\mathsfit}{bold}{\encodingdefault}{\sfdefault}{bx}{n}

\usepackage{hyperref}
\usepackage{xspace}
\usepackage{url}
\usepackage{graphicx}
\usepackage{booktabs}
\usepackage{siunitx}
\usepackage{adjustbox}
\usepackage{makecell}
\usepackage{tabularx}
\usepackage{threeparttable}
\usepackage{enumitem}
\usepackage{amsmath,amssymb}
\newcommand{\name}{PINATA\xspace}

\title{Evolutionary Informed Translation from Protein to Nucleic Acid Sequence}

\author{Anna Hart \\
  University of Illinois Urbana-Champaign \\
  Siebel School of Computing and Data Science \\
  DOE Center for Advanced Bioenergy and Bioproducts Innovation, University of Illinois Urbana-Champaign\\
  \texttt{anna19@illinois.edu}
  \And
  Mohammed El-Kebir \\
  University of Illinois Urbana-Champaign \\
  Siebel School of Computing and Data Science \\
  \AND
  Huimin Zhao \\
  University of Illinois Urbana-Champaign \\
  Chemical and Biomolecular Engineering \\
  DOE Center for Advanced Bioenergy and Bioproducts Innovation, University of Illinois Urbana-Champaign\\
  \And
  Heng Ji \\
  University of Illinois Urbana-Champaign \\
  Siebel School of Computing and Data Science \\
  DOE Center for Advanced Bioenergy and Bioproducts Innovation, University of Illinois Urbana-Champaign\\
  \texttt{hengji@illinois.edu}
}

\iclrfinalcopy 
\begin{document}

\maketitle

\begin{abstract}

The choice of nucleic acid sequence for a given protein strongly impacts function and is species-specific. Current protein-to-nucleic-acid translation models treat sequences in isolation, often over-favoring frequently used codons. To generate more natural nucleic acid sequences, we introduce \name (Phylogenetically Informed protein to Nucleic Acid Translation with sequence Alignment), a model that leverages sequences from related species encoded in multiple sequence alignments (MSAs) to guide prediction. Our architecture leverages evolutionary context and learns species-and protein specific context to make biologically-informed predictions. Evaluated across 61,934 gene trees and 882 species, \name generates sequences that mirror the naturally-occurring sequence substantially better than existing language models. \name leverages MSA data to improve exact native codon recovery by 17.95 percentage points over existing language models, effectively preserves natural rare-codon nuances, and more closely matches native sequence properties.  \name maintains strong predictive performance even with limited evolutionary data. While previous methods largely treated sequences from different species in isolation, our method leverages the evolutionary relationships between nucleic acids and proteins across species to create a more biologically-informed translation. We demonstrate that leveraging rich evolutionary data through an biologically-designed method strongly aids in translating protein sequences to nucleic acid sequences across species. We will make data and code publicly available upon publication.

\end{abstract}

\section{Introduction}

Biological information is encoded through large molecular sequences; these sequences and their interactions create biological language. Two fundamental sequence representations are nucleic acid sequences (DNA and RNA), which store and transmit key information using four nucleotide bases, and proteins, which are composed of 20 building blocks called amino acids and carry out a wide range of functions. These sequence representations can be viewed as distinct biological languages connected by the genetic code, which maps nucleotide triples, called codons, to amino acids during protein synthesis. While each codon specifies a single amino acid, the reverse translation from proteins to nucleic acids is not deterministic. There are $4^3= 64$ nucleic acid codons but only $20$ amino acids, so most amino acids are encoded by multiple synonymous codons, and an exponential number of synonymous nucleic acid sequences code for the same protein.  This choice of nucleic acid sequence has consequential biological effects \citep{kudla2009coding, boel2016codon, varenne1984translation} and depends on both the properties of the protein and on the species (e.g., dogs, mice, \textit{E.\ coli}). This intricate relationship means that even the same protein takes on a different nucleic acid sequence depending on the species in which it is expressed. This is especially relevant in bioengineering, where genes are frequently transferred across species \citep{demain2009production, luo2015engineered, huo2019heterologous}. Generating a native-like sequence could provide a starting point for multi-objective gene design that can be further optimized for specific qualities desired in bioengineering applications such as protein expression and solubility. 
We frame the problem as a translation spanning multiple axes of biological language: from protein language to nucleic acid language, across the languages of many species. This multi-faceted translation requires reasoning jointly over protein sequence, nucleic acid sequence, and species identity, offering a useful lens on multi-modal biological language modeling. 
With this motivation, we pose the central problem of our paper: how do we translate from protein language to nucleic acid language across species? 

\begin{figure}
    \centering
    \includegraphics[width=1\linewidth]{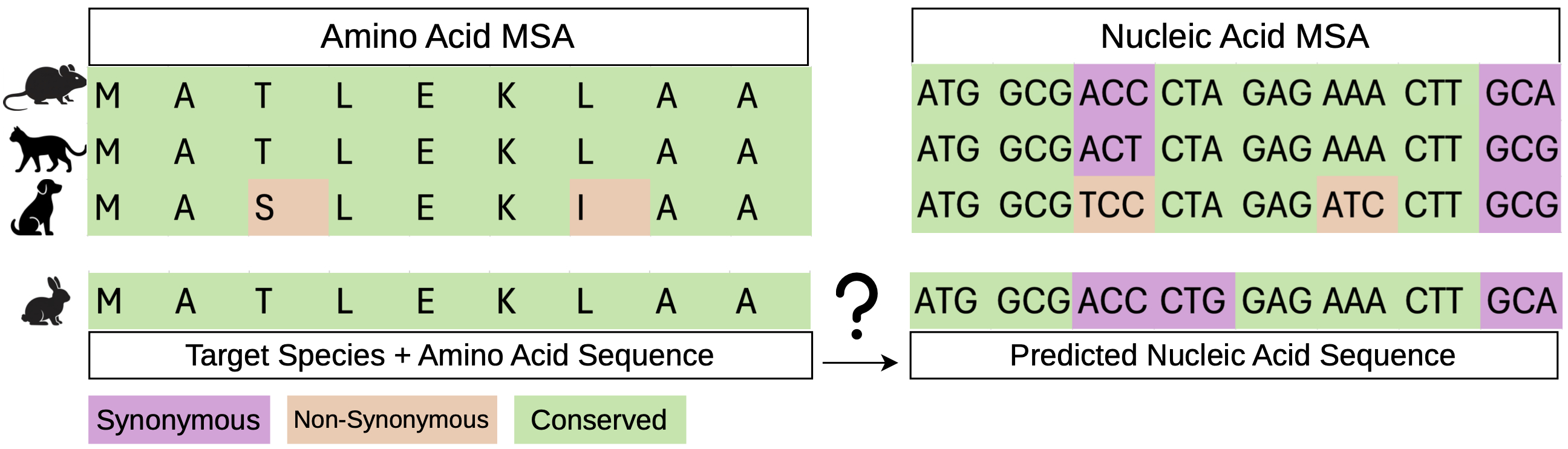}
    \caption{\textbf{Evolutionary Informed Translation from Protein to Nucleic Acid Sequence}. Protein sequences of multiple species (mouse, cat, dog, and rabbit) are aligned in the amino acid MSA (left). Each protein sequence is made up of amino acids (denoted with 1-letter codes) and each nucleic acid sequence is made up of a sequence of codons (denoted with 3-nucleotides per codon). The known nucleic acid sequences (mouse, cat, dog) are aligned in the nucleic acid MSA (right), by following the amino acid MSA. For the target rabbit protein sequence (MATLEKLAA), we predict the corresponding rabbit nucleic acid sequence (right). Conserved positions (green) are the same in both amino acid and nucleic acid choice. Synonymous (purple) positions are the same for the amino acid yet differ for the nucleic acid. Non-synonymous (orange) positions differ for both the amino acid and the nucleic acid. Our goal is to predict the nucleic acid sequence for the target protein and species by leveraging information in the MSAs.}
    \label{fig:intro}
\end{figure}

A key aspect of the problem is how nucleic acid languages across different species are related through evolution. 
The relatedness between nucleic acid sequences provides a key biological signal for understanding and recovering the natural sequence distributions across species. 
While each species has its own version of nucleic acid language, 
related species - such as chimps and humans - share high similarity in their nucleic acid sequences, with differences between the languages accumulating over time and giving rise to the species differences we see today \citep{dobzhansky1982genetics}. 
In particular, similarities between the languages can point to key functionality preserved for survival while differences can point to species-dependent features. 
To represent these evolutionary changes across species, biologists use multiple sequence alignments (MSAs) \citep{feng1987progressive}, where each row is a species and each column represents how a given position in the sequence changes across species (Figure \ref{fig:intro}).
Specifically, paired MSAs, composed of protein and corresponding nucleic acid sequences, are constructed for homologous sequences (related sequences with the same or similar amino acid sequence from different species).
For a given position in homologous sequences, three possibilities occur. First, the position can be conserved: the amino acid and the nucleic acid codon in both sequences are the same. Second, the position can have a synonymous change: the amino acid is the same but the codon is different. Third, the position can have a non-synonymous change: both the amino acid and the codon differ.

Critically, this evolutionary context is available for many proteins of interest. When translating a protein to a nucleic acid sequence of a particular species, we can leverage the homologous sequences from related species. This data can easily be found in databases such as Ensembl \citep{yates2026ensembl} or constructed with existing bioinformatics methods \citep{katoh2013mafft}. Certain species are more well-studied than others; leveraging MSA information enables context from well-studied species to be used when making predictions regarding data-scarce species. Current models for translating protein to nucleic acid sequences do not leverage MSA information nor do they directly account for evolution, treating species largely in isolation with the goal that a large enough model will learn patterns out of millions of sequences \citep{heuschkel2026advancing, fallahpour2025codontransformer}. Instead of modeling the problem as an evolutionary-informed translation across species, they represent the species as a context token, without capturing the evolutionary relationships between protein and nucleic acid sequences across species. Since these models are trained with the objective of selecting the correct codon, they tend to favor high-frequency codons, deviating from the species' natural sequence distribution.

In our work, we directly leverage the evolutionary relatedness between the nucleic acid languages of different species. By deeply integrating our model with the biological evolutionary process, we recover sequences much closer to the native (found-in-nature) sequence. 
We introduce \name,  a method that efficiently injects evolutionary information at every position in the amino acid sequence and leverages attention to allow each token to attend to the context of the surrounding amino acids, the surrounding evolutionary context, and the species. This enables our model to uniquely adhere to the rich information encoded in homologous sequences while adapting to protein and species specific context. Our model creates more natural (closer to the native sequence), species-specific nucleic acid sequences across a wide variety of proteins and over 800 species. 
\name offers three key contributions: 
\begin{itemize}[noitemsep, topsep=0pt]
    \item We leverage evolutionary MSA data from related species as key biological information for translating between protein and species-dependent nucleic acid sequences
    \item We deeply integrate the evolutionary information throughout our model and allow the model to learn species and protein specific context, creating a biologically-aligned design
    \item \name generates sequences that are substantially more similar to the native sequences than previous work, with a 17.95pp higher recovery of native codons compared to existing language models. We demonstrate that \name's improvements persist even when evolutionary data is scarce.
\end{itemize}
\section{Related Work}
Several language modeling approaches have been developed to convert a protein to a nucleic acid sequence of a given species. CodonTransformer \citep{fallahpour2025codontransformer} utilizes an encoder-only architecture trained with a masked-language-modeling (MLM) objective. By adding a species encoding to amino acid-codon pair tokens, the model predicts masked tokens as the most likely codon given the amino acid sequence and species. SynCodonLM \citep{heuschkel2026advancing} builds upon this approach by explicitly restricting the model to only predict synonymous codons. ppLMCo \citep{pathak2026pplm} improves the efficiency of codon prediction with a linear layer on top of a pre-trained encoder. Very recently, CoFormer \citep{kim2026enhancing} integrates transformers and CNNs. GEMORNA \citep{zhang2025deep} is trained on mammalian genes, using an encoder-decoder architecture to predict the corresponding nucleic acid sequence, additionally generating untranslated regions outside of the coding sequence to influence expression. There are additional papers \citep{li2025deep, rajbanshi2025codongpt, lin2026rnarl} that seek to optimize nucleic acid sequences for specific properties. These methods model species independently, without representing the evolutionary relationships between them. In our work, we demonstrate that evolution-based modeling renders recovery of native sequences. 

A few models incorporate limited amounts of evolutionary data in their approach. \citet{sidi2025predicting}, OrthologTransformer \citep{akiyama2026cross}, and CodonT5 \citep{babjac2024codont5} give a single homologous protein as context. This does not, however, account for the large amount of evolutionary information encoded in dozens or hundreds of homologous sequences in the MSA. DeepCodon \citep{han2025deepcodon} is trained solely on Enterobacteriaceae sequences, but uses homologous sequences to derive a binary signal for whether or not a position contains a rare codon. CodonTranslator \citep{chen2025codontranslator} is a preprint which includes information about the relatedness between species in the species token; however, it does not include information about sequence relatedness through MSAs. Outside of our problem, there are modeling approaches for incorporating evolutionary information, such as encoding MSAs \citep{rao2021msa}, predicting protein structure \citep{jumper2021highly}, predicting variant effects \citep{riesselman2018deep, frazer2021disease}, and predicting protein fitness \citep{notin2022tranception}; however, these methods do not translate from protein to nucleic acid languages across species. 

\section{Methods}
\begin{figure}
    \centering
    \includegraphics[width=1\linewidth]{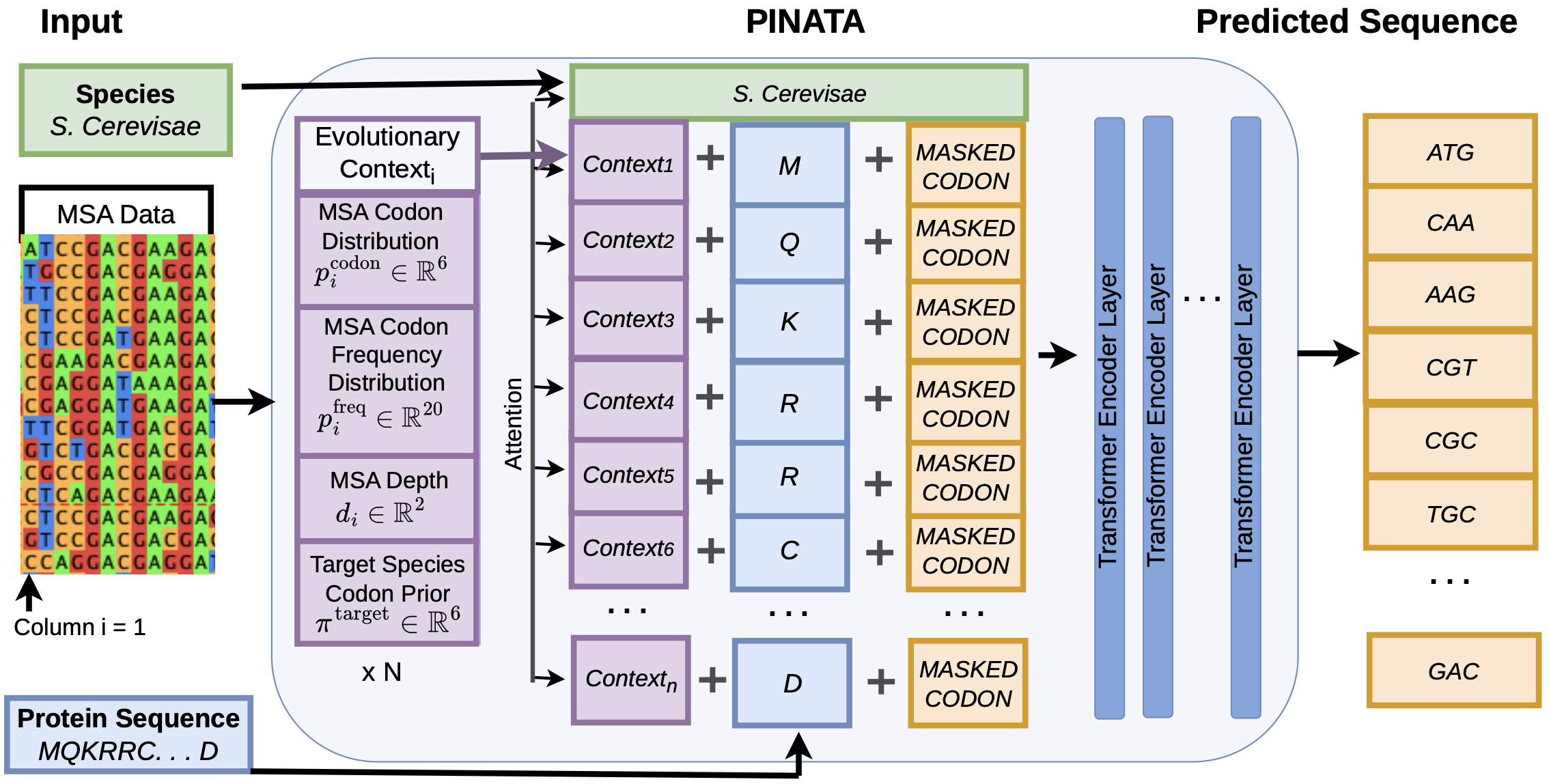}
    \caption{Overview of \name. \name extracts an evolutionary context vector from every position in the MSA. The context vector at every position is added to the amino acid and codon tokens. Attention occurs between the species token and the context-amino acid-codon tokens throughout a multi-layer transformer encoder. \name then decodes a synonymous nucleic acid sequence, taking into account the evolutionary context, protein context, and species-specific patterns. MSA Data visualized with JalView \citep{Procter2021}}
    \label{fig:modeloverview}
\end{figure}

We design a bio-inspired model to deeply integrate evolutionary information into the translation between an amino acid sequence to the nucleic acid sequence across species. Our model is tailored to leverage the relatedness of sequences across species while enabling the model to capture nuance within the species and protein-specific context. An overview of our model is shown in Figure \ref{fig:modeloverview}.

\subsection{Evolutionary Context}\label{sec:evolutionary}
We first design how to extract key information from the evolutionary context of the sequences from related species. To do this, we leverage MSA data encoding how nucleic acid sequences for a given protein change across species at each position of the protein. Details of this data are described in Section \ref{sec:data}. 

Each MSA consists of tens to hundreds of homologous sequences, amounting to a structure of size NxM, where N is the length of the target protein and M is the number of homologous sequences. We design a module to extract the most relevant biological information from this data into an efficient structure of 34 x N, where each column is a context vector of length 34. 

There are two main pieces of evolutionary information that we extract at each position. First, for each target amino acid, we calculate the distribution of corresponding codons at that position in each homologous sequence. Since there is often significant similarity between nucleic acid sequences of related species, the corresponding homologous codons provides a strong signal. We define this information as MSA Codon Distribution, $\mathbf{p}^{\mathrm{codon}}_i
=
\left(
p^{\mathrm{codon}}_{i,1},
\ldots,
p^{\mathrm{codon}}_{i,6}
\right)
\in\mathbb{R}^{6}$ where $p^{\mathrm{codon}}_{i,n}$ represents the frequency of the $n^{th}$ codon corresponding to the target amino acid $a_i$ in column $i$. Each value is calculated across all non-target species in the MSA with the target amino acid at position $i$. As there are up to 6 codons per amino acid, the MSA Codon Distribution is in $\mathbb{R}^{6}$. 

Each species has its own preferences for specific codons \citep{grantham1980codon}. As a result, when a given amino acid position should be encoded by a rare codon, the codon actually used at that position in two different species will depend on the relative frequency, or preference, of that codon in each species' genome. To extract this codon frequency signal, we first compute the species-specific codon frequencies for all codons $c$ and species $s$ as \(\pi_{s,c} = N_{s,c}/\sum_{c' \in \mathcal{C}(a)}N_{s,c'}\), where $a$ is the amino acid encoded by c, \(N_{s,c}\) is the number of occurrences of codon \(c\) in species \(s\), and \(\mathcal{C}(a)\) is the set of synonymous codons encoding \(a\). Importantly, we only use training sequences when constructing this data. Then, for each position $i$ in our target protein, we construct a histogram of the relative frequencies across corresponding codons. We define this histogram as the MSA Codon Frequency Distribution, $\mathbf{p}^{\mathrm{freq}}_i
=
\left(
p^{\mathrm{freq}}_{i,0},
\ldots,
p^{\mathrm{freq}}_{i,19}
\right)
\in\mathbb{R}^{20}$. To construct this histogram, for each synonymous codon $j$ at position $i$, we obtain the relative frequency \(\pi_{s_j,c_{ij}}\) of its observed codon \(c_{ij}\) in its species \(s_j\). These values are assigned to 20 bins, of width 0.05 and normalized over the comparable homologous sequences. Thus, the value at each bin $b$ for position $i$ can be calculated as  \( p^{\mathrm{freq}}_{i,b}=\frac{1}{|H_i|}\sum_{j\in H_i}\mathbb{I}\!\left[\min\!\left(\left\lfloor20\pi_{s_j,c_{ij}}\right\rfloor,19\right)=b\right],\quad b\in\{0,\ldots,19\}\), where \(H_i\) is the set of homologous sequences that contain a codon at position \(i\) encoding the same amino acid as the target.
\newline \newline
We include two additional values as context. First, we recognize that not every target amino acid position has an equivalent number of synonymous codons in the homologous sequences (for example, some homologous sequences may have a gap or a different amino acid). We therefore include two values for how much support there is at a given position, $i$. We express this as $\mathbf{d}_i
=
\left[
\frac{\log(1+n_i)}
     {\log\!\left(1+\max(1,N)\right)},
\;
\frac{n_i}{\max(1,N)}
\right]
\in\mathbb{R}^{2}$, where $N$ is the total number of homologous sequences in the MSA after excluding the target species and $n_i$ is the number of homologous sequences matching the target amino acid at that position. 

Finally, we add species-specific context into each position to avoid over-fitting to homologous sequences. At each amino acid position, we provide the target species codon prior, \(\pi_{c} = N_{c}/\sum_{c' \in \mathcal{C}(a)}N_{c'}\) where $c$ is a given codon encoding amino acid $a$, $N_c$ is the total number of codons $c$ across all training sequences, and $\sum_{c' \in \mathcal{C}(a)}N_{c'}$ is the total number of codons encoding amino acid $a$ across all training sequences of the target species. This is not evolutionary information and provides information regarding codon preferences to the model.

Placing all these contexts together, $p_i^{\mathrm{codon}}$, $p_i^{\mathrm{freq}}$, $d_i$, and $\pi_c$ gives a 34-dimensional vector at each amino acid position $i$. We then project this context to the 768 dimensions of the amino acid and codon embeddings through a linear layer, followed by layer-normalization, and GELU \citep{hendrycks2016gaussian} activation.

\subsection{Architecture}
Our model efficiently injects the evolutionary context as a 34-feature vector described in Section \ref{sec:evolutionary} at every amino acid position. We add the projected biological context, the amino acid, and the codon embeddings together. This directly fuses the most relevant evolutionary context into each amino-acid/codon embedding. 

We leverage attention to enable the model to learn nuanced protein and species-specific patterns. Attention within each transformer layer shares information between each joint evolutionary-context/amino-acid/codon embedding with the species embedding prepended to the sequence; allowing each position to account for its own and surrounding evolutionary context, surrounding amino acids, and species. The species vocabulary contains 882 species plus an unknown species token. 

Our model uses a fixed sinusoidal positional encoding and passes the tokens through an encoder-only Transformer \citep{vaswani2017attention} trained with BERT-style \citep{devlin2019bert} masked language modeling (MLM). The model has 13 layers, a hidden dimension of 768, 12 self-attention heads of dimension 64, and a 2,944-dimensional GELU \citep{hendrycks2016gaussian} feed-forward layer. Each Transformer block uses pre-layer normalization and a dropout rate of 0.1. The output head produces six codon-slot logits. Slots not corresponding to a valid synonymous codon for the input amino acid are masked before decoding. The model contains approximately 90M parameters. Additional details are in Appendix \ref{app:data}. 

We limit each model input to a sequence length of 2,048. During training, longer sequences are truncated to the first 2,048 positions; during inference, they are divided into consecutive non-overlapping chunks of at most 2,048 positions, decoded independently with the species token repeated for each chunk, and concatenated to produce the sequence.

\subsubsection{Training}
During the first four epochs, we train the model to maximize the probability of native codons at masked positions given the target amino-acid sequence, the partially masked codon sequence, the evolutionary context, and the target species. We randomly replace approximately \(15\%\) of eligible codon positions with a universal mask token and use the MLM cross-entropy loss
\[
\mathcal{L}_{\mathrm{naturalCE}}
=
-\frac{1}{|\mathcal{M}|}
\sum_{i\in\mathcal{M}}
\log
p_{\theta}
\left(
c_i
\mid
\mathbf{a},
\widetilde{\mathbf{c}},
\mathbf{X}^{\mathrm{context}},
s
\right),
\]
where \(\mathcal{M}\) is the set of masked positions, \(c_i\) is the native ground-truth codon at position \(i\), \(\mathbf{a}\) is the target amino-acid sequence, \(\widetilde{\mathbf{c}}\) is the codon sequence with positions in \(\mathcal{M}\) replaced by mask tokens, \(\mathbf{X}^{\mathrm{context}}\) is the position-specific evolutionary context matrix, and \(s\) is the target species. 

In the 5th epoch, we add an additional component to the loss, $\mathcal{L}_{\mathrm{property}}$ to maximize how well the predicted sequence follows the natural properties of the ground truth sequence. We define $\mathcal{L}_{\mathrm{property}}$ as a combination of terms which penalize according to the difference between properties of the predicted and natural sequence. We define $\mathcal{L}_{\mathrm{property}} = \frac{1}{4}
\left[
\frac{|\Delta\mathrm{GC}|}{5}
+
\frac{|\Delta\mathrm{CFD}|}{5}
+
\frac{|\Delta\mathrm{CSI}|}{0.1}
+
\frac{\operatorname{softDTW}_{\mathrm{norm}}}{10}
\right]$ Here, $\Delta{GC}$ is the difference of the percentage of G and C content across all codons between the native and predicted sequence. $\Delta{CFD}$ \citep{fallahpour2025codontransformer} is the difference in the amounts of rare codons, $\Delta{CSI}$ is the codon similarity index, measuring how much the sequence favors codons with a high frequency \citep{sabath2012evolution}, and normalized soft-DTW divergence \citep{cuturi2017soft, blondel2021differentiable} is a proxy for DTW distance \citep{giorgino2009computing} which measures how well the predicted sequence's codon frequency distribution mirrors that of the native sequence.

We combine the MLM loss with the property loss to create the final-epoch loss $\mathcal{L}_\mathrm{finalEpoch} = \mathcal{L}_{\mathrm{property}} + \mathcal{L}_{\mathrm{naturalCE}}$. We use a 100\% mask in the 5th epoch to better represent the downstream task and fine-tuned for up to one additional epoch and selected among quarter-epoch validation checkpoints.

\section{Results}

\subsection{Data Collection}\label{sec:data}

We collect MSAs from aligned protein gene trees in Ensembl Compara release 116 and Ensembl Genomes release 63 \citep{yates2026ensembl}, encompassing Fungi (EFGT0), Metazoa (EMGT0), Vertebrates (ENSGT), Plants (EPIGT), and Protists (EPrGT). Notably, these gene trees each comprise of loosely related species (e.g., all fungi or all plants). We combine all of the MSAs from these gene trees into our dataset. Thus, our model works for many different types of species, and the MSA for each prediction is from a grouping of loosely related species. 

We select one representative sequence per species within each gene tree and perform quality control, with details in the Appendix \ref{app:conservation}. We retain trees containing at least 25 valid sequences from at least 25 distinct species. The resulting dataset consists of 61,934 MSAs and 5,933,724 protein-coding sequences across 882 distinct species. Dataset composition is reported in Table \ref{tab:data}. 

To limit leakage between training and evaluation sets, we cluster all protein sequences using MMseqs2 \citep{steinegger2017mmseqs2} at thresholds of at least 80\% amino-acid identity and 80\% coverage in both the query and target directions. Gene trees connected through any sequence cluster are treated as a connected component. Components are assigned to training, standard validation, and standard test sets, targeting a 90\%/5\%/5\% distribution by sequence count. For the standard set, no member of the target MSA appears during training, and members of the target MSA from non-target species provide evolutionary context at inference.

We also construct a within-tree test set composed of proteins with homologous sequences present during training by holding out approximately 5\% of sequences from train-set MSAs. The purpose of this test set is to measure how our model compares even when all models have seen the homologous sequences during training. We randomly generate 3 different splits of the dataset.

\begin{table*}[t]
  \centering
  \small
  \caption{Dataset distribution and synonymous-codon conservation}
  \label{tab:data}
  \setlength{\tabcolsep}{5pt}
  \begin{tabular}{@{}lrrrr@{}}
    \toprule
    Division & Gene trees & \makecell{Sequences (\%)} & \makecell{Observed\\conservation (\%)} & \makecell{Null\\conservation (\%)} \\
    \midrule
      Fungi & 16{,}240 & 1{,}650{,}655 (27.82\%) & 39.98 & 34.27 \\
      Metazoa & 8{,}141 & 422{,}104 (7.11\%) & 41.63 & 33.90 \\
      Vertebrates & 18{,}345 & 2{,}446{,}863 (41.24\%) & 60.70 & 35.68 \\
      Plants & 14{,}242 & 1{,}138{,}013 (19.18\%) & 57.59 & 33.91 \\
      Protists & 4{,}966 & 276{,}089 (4.65\%) & 47.06 & 37.62 \\
      \midrule
      \textbf{Total} & 61{,}934 & 5{,}933{,}724 (100.00\%) & 50.08 & 34.91 \\
    \bottomrule
  \end{tabular}
\end{table*}

We demonstrate that MSAs from loosely related species encode rich biological information due to conservation. We measure synonymous-codon conservation by comparing codon choices among homologous residues encoding the same amino acid and contrast the observed agreement with the agreement expected from species-specific codon usage. We find that across every division of our dataset, codon conservation produces a strong signal beyond what would be expected by random chance, demonstrated in Table \ref{tab:data}.

To evaluate out-of-distribution generalization, we construct 1,000 bacterial
codon MSAs from GTDB release R232 \citep{parks2026gtdb}. We select 500
high-quality representative genomes spanning 10 bacterial phyla, predict
protein-coding genes with Prodigal \citep{hyatt2010prodigal}, and infer
single-copy orthogroups separately within each phylum using OrthoFinder
\citep{emms2019orthofinder}. Orthogroups containing at least 25 species are
aligned at the amino-acid level with MAFFT \citep{katoh2013mafft} and
back-translated to codon alignments using the corresponding coding sequences.
We select one target from each of 1,000 distinct orthogroups and exclude
targets with at least 80\% sequence identity and 80\% bidirectional coverage to any training protein using MMSeqs2 \citep{steinegger2017mmseqs2} from the training.

\subsection{Evaluation Metrics}
We evaluate how well our generated codon sequences represent the distribution found in the native sequence through the following metrics: 
\begin{itemize}
    \item Exact Codon Match: The percentage of codon positions where the predicted codon is identical to the native codon
    \item GC Content: The percentage of G and C nucleotides in the predicted codon sequence
    \item Codon Similarity Index (CSI) \citep{sabath2012evolution}: The geometric mean of positive relative codon adaptiveness values. We use the frequency of codons by species  in our full dataset as the ground-truth frequencies. CSI demonstrates how the sequence favors codons that are more frequently used by the species.
    \item Codon Frequency Deviation (CFD): Calculated as in \citet{fallahpour2025codontransformer}: the percentage of predicted codons with relative adaptiveness below 0.3.
    \item Minimum Free Energy (MFE): The predicted mRNA folding energy, calculated using ViennaRNA \citep{lorenz2011viennarna}. Due to the computational time needed to compute MFE, MFE was computed on a random subset of 50,000 sequences with a sequence length limit of 2,048. The same subset was used across all models. 
    \item DTW Distance: As in \citet{fallahpour2025codontransformer}, we follow the algorithm in \citet{giorgino2009computing}, the dynamic time warping distance between the predicted and native sliding window of \%MinMax \citep{clarke2008rare} codon adaptation profiles (18-codon window). DTW Distance measures how well the predicted sequence's codon frequency distribution mirrors that of the native sequence. 
\end{itemize}

To demonstrate how well the predictive sequence matches the properties of the native sequence, we present GC content, CSI, CFD, and MFE as the sequence-level absolute difference between the predicted and native sequence, macro-averaged over sequences. 

\subsection{Model Performance}
In our experiments, we answer the following questions: (1) Does \name generate more native-like sequences than state-of-the-art models? (2) What is the cause behind \name's improvements? (3) Do \name 's improvements persist even when all models have seen the homologous sequences during training, (4) How does \name perform when the amount of MSA data is scarce, and (5) How does \name perform out-of-distribution? 

We compare our model against two state-of-the art language modeling methods for species-dependent conversion of protein sequences to nucleic acid sequences: CodonTransformer \citep{fallahpour2025codontransformer} and SynCodonLM \citep{heuschkel2026advancing}. For a controlled comparison, we train both methods from scratch on the three same training splits used by our model, using the same five-epoch training budget and selecting checkpoints using the standard dataset validation performance. For SynCodonLM, we are only able to evaluate 50,000 sequences with length no greater than 2,048 on the standard dataset and 10,000 sequences with length no greater than 2,048 on the within-tree set per run due to the computational time required by its inference: these results are included in the main tables with a comparison against all models on the exact subset of sequences shown in the Appendix \ref{app:syncodonset}.  We adhere as closely as possible to each method's original implementation and training procedure, with complete details provided in the Appendix \ref{app:comparators}. 

To answer question 1, we test our model on the set of held-out gene trees. \name outperforms the leading language model at predicting native-like sequences across all metrics, including a 17.95 percentage point higher codon match and 68.0\% lower DTW (Table \ref{tab:wholetree}). These results demonstrate that \name generates sequences with significantly higher similarity to the native sequence than previous language models. 

To demonstrate evolutionary context as the source of \name's improvement (question 2), we ablate all MSA-derived features (the homolog codon distribution, homolog frequency histogram, and MSA depth) at both training and inference. Removing MSA data degrades performance on all metrics (Table \ref{tab:wholetree}), confirming that evolutionary context is central to the model. We implement a simple heuristic that, at each position, selects the most frequently observed synonymous codon among homologs encoding the target amino acid, falling back to the target species' training-derived codon frequency prior when no such homolog exists. This heuristic, shown as MSA signal only (no learning) in the table, alone outperforms SOTA language models on every metric, improving codon match by 13.02pp. Learning nuanced context contributes further gains: \name adds a further 4.93pp gain in codon match and improvements on every metric over the heuristic, demonstrating the importance of leveraging MSA information with context-specific learning. 

To show \name 's improvement even when all models have seen the homologous sequences during training (question 3), we use a within-tree test-set. We hold out approximately 5\% of sequences from each training-set gene tree for the test set and place the homologous sequences in the training set. All models see the homologous sequences during training. In this setting, \name recovers a substantially more natural distribution than leading language models, shown in Table \ref{tab:withintree} with improvements similar to those in the standard data set, showing that our methodology for incorporating evolutionary context from homologous sequences at inference substantially boosts recovery of the native nucleic acid sequence.

\begin{table*}[t]
  \centering
  \small
  \begin{adjustbox}{max width=\textwidth}
  \begin{threeparttable}
    \caption{Model Performance (Standard Dataset)}
    \label{tab:wholetree}
    \setlength{\tabcolsep}{4.5pt}
    \begin{tabular}{@{}l *{6}{c}@{}}
      \toprule
      Model & \makecell{Codon match\\(\%) $\uparrow$} & \makecell{$|\Delta \mathrm{GC}|$\\(pp) $\downarrow$} & \makecell{$|\Delta \mathrm{CFD}|$\\(pp) $\downarrow$} & \makecell{$|\Delta \mathrm{CSI}|$\\$\downarrow$} & \makecell{Normalized\\DTW $\downarrow$} & \makecell{$|\Delta \mathrm{MFE}|$\\(kcal/mol) $\downarrow$} \\
      \midrule
      \name (Ours) & \textbf{67.94 $\pm$ 0.06} & \textbf{1.426 $\pm$ 0.004} & \textbf{0.962 $\pm$ 0.003} & \textbf{0.0209 $\pm$ 0.0001} & \textbf{4.538 $\pm$ 0.004} & \textbf{17.95 $\pm$ 0.12} \\
      CodonTransformer & 49.99 $\pm$ 0.09 & 6.859 $\pm$ 0.478 & 3.218 $\pm$ 0.076 & 0.1189 $\pm$ 0.0003 & 14.202 $\pm$ 0.423 & 73.39 $\pm$ 4.78 \\
      SynCodonLM & 43.21 $\pm$ 1.30 & 11.055 $\pm$ 1.377 & 3.812 $\pm$ 0.353 & 0.1237 $\pm$ 0.0152 & 19.089 $\pm$ 2.492 & 104.00 $\pm$ 19.95 \\
      \name (no MSA ablation) & 50.35 $\pm$ 0.08 & 2.843 $\pm$ 0.039 & 1.502 $\pm$ 0.010 & 0.0407 $\pm$ 0.0004 & 6.387 $\pm$ 0.007 & 31.36 $\pm$ 0.89 \\
      MSA signal only (no learning) & 63.01 $\pm$ 0.01 & 2.698 $\pm$ 0.001 & 1.436 $\pm$ 0.006 & 0.0353 $\pm$ 0.0001 & 5.361 $\pm$ 0.009 & 24.91 $\pm$ 0.12 \\
      \bottomrule
    \end{tabular}
    \begin{tablenotes}[flushleft]
      \footnotesize
      \item Values are mean $\pm$ sample standard deviation over 3 runs. Metrics were calculated over approximately 296,000 sequences and MFE is evaluated on each run's shared frozen 50,000-sequence cohort. SynCodonLM calculates its evaluation metrics over a subset of 50,000 sequences due to computational expense.
    \end{tablenotes}
  \end{threeparttable}
  \end{adjustbox}
\end{table*}

\begin{table*}[t]
  \centering
  \small
  \begin{adjustbox}{max width=\textwidth}
  \begin{threeparttable}
    \caption{Model Performance (Within-tree dataset)}
    \label{tab:withintree}
    \setlength{\tabcolsep}{4.5pt}
    \begin{tabular}{@{}l *{6}{c}@{}}
      \toprule
      Model & \makecell{Codon match\\(\%) $\uparrow$} & \makecell{$|\Delta \mathrm{GC}|$\\(pp) $\downarrow$} & \makecell{$|\Delta \mathrm{CFD}|$\\(pp) $\downarrow$} & \makecell{$|\Delta \mathrm{CSI}|$\\$\downarrow$} & \makecell{Normalized\\DTW $\downarrow$} & \makecell{$|\Delta \mathrm{MFE}|$\\(kcal/mol) $\downarrow$} \\
      \midrule
      \name (Ours) & \textbf{65.77 $\pm$ 0.06} & \textbf{1.448 $\pm$ 0.016} & \textbf{0.964 $\pm$ 0.008} & \textbf{0.0217 $\pm$ 0.0000} & \textbf{4.600 $\pm$ 0.007} & \textbf{21.34 $\pm$ 0.17} \\
      CodonTransformer & 49.97 $\pm$ 0.48 & 6.731 $\pm$ 0.602 & 3.197 $\pm$ 0.059 & 0.1110 $\pm$ 0.0005 & 12.366 $\pm$ 0.427 & 86.70 $\pm$ 6.58 \\
      SynCodonLM & 42.78 $\pm$ 1.04 & 11.153 $\pm$ 1.431 & 4.073 $\pm$ 0.383 & 0.1247 $\pm$ 0.0131 & 18.167 $\pm$ 2.335 & 122.97 $\pm$ 21.78 \\
      \name (no MSA ablation) & 50.63 $\pm$ 0.11 & 2.616 $\pm$ 0.022 & 1.395 $\pm$ 0.009 & 0.0382 $\pm$ 0.0002 & 5.954 $\pm$ 0.013 & 34.82 $\pm$ 1.51 \\
      MSA signal only (no learning) & 58.49 $\pm$ 0.01 & 3.737 $\pm$ 0.011 & 2.081 $\pm$ 0.006 & 0.0487 $\pm$ 0.0001 & 6.348 $\pm$ 0.008 & 37.24 $\pm$ 0.08 \\
      \bottomrule
    \end{tabular}
    \begin{tablenotes}[flushleft]
      \footnotesize
      \item Values are mean $\pm$ sample standard deviation over 3 runs. Metrics were calculated over approximately 269,000 sequences and MFE is evaluated on each run's shared frozen 50,000-sequence cohort. SynCodonLM calculates its evaluation metrics over a subset of 10,000 sequences due to computational expense.
    \end{tablenotes}
  \end{threeparttable}
  \end{adjustbox}
\end{table*}

We next evaluate \name when MSA data is scarce (question 4), progressively removing the 10\%, 25\%, 50\%, 75\%, 90\%, and 100\% most closely related homologs (Table \ref{tab:msa-depth-ablation}). \name outperforms the no-MSA ablation with up to 90\% of homologs removed. With the MSA removed entirely, it falls below that ablation, since it was never trained without evolutionary context.

\begin{table*}[t]
  \centering
  \small
  \begin{adjustbox}{max width=\textwidth}
  \begin{threeparttable}
    \caption{Model Performance with Reduced MSA Context}
    \label{tab:msa-depth-ablation}
    \setlength{\tabcolsep}{6pt}
    \begin{tabular}{@{}lrrrrrr@{}}
      \toprule
      \makecell{Closest homologs\\removed (\%)} & \makecell{Codon match\\(\%) $\uparrow$} & \makecell{$|\Delta \mathrm{GC}|$\\(pp) $\downarrow$} & \makecell{$|\Delta \mathrm{CFD}|$\\(pp) $\downarrow$} & \makecell{$|\Delta \mathrm{CSI}|$\\$\downarrow$} & \makecell{Normalized\\DTW $\downarrow$} & \makecell{$|\Delta \mathrm{MFE}|$\\(kcal/mol) $\downarrow$} \\
      \midrule
      10 & 62.86 & 1.692 & 1.100 & 0.0240 & 4.976 & 20.63 \\
      25 & 60.12 & 1.856 & 1.178 & 0.0255 & 5.231 & 22.16 \\
      50 & 57.66 & 2.014 & 1.267 & 0.0270 & 5.446 & 23.33 \\
      75 & 54.48 & 2.274 & 1.386 & 0.0294 & 5.729 & 25.60 \\
      90 & 52.00 & 2.510 & 1.507 & 0.0318 & 6.010 & 27.61 \\
      100 & 45.89 & 4.161 & 2.243 & 0.0436 & 7.784 & 39.68 \\
      \midrule
      No-MSA ablation& 50.38 & 2.828 & 1.504 & 0.0406 & 6.383 & 30.96 \\
      \bottomrule
    \end{tabular}
    \begin{tablenotes}[flushleft]
      \footnotesize
      \item Homologs removed in order of increasing target-to-homolog patristic distance, so the phylogenetically closest sequences were removed first. Metrics were calculated over 296,703 sequences and MFE uses the same frozen cohort of 50,000 sequences across one run
    \end{tablenotes}
  \end{threeparttable}
  \end{adjustbox}
\end{table*}

In Table \ref{tab:bacteria}, we demonstrate how \name performs out of distribution (question 5) with a bacteria dataset. No models were trained on bacteria. To minimize information about the unseen species, we gave only a uniform prior instead of the target species codon frequency prior. For the 1,000 bacterial sequences predicted, \name exhibited the best performance. Inclusion of MSA information improves model accuracy even out of distribution. 

\begin{table*}[t]
  \centering
  \small
  \begin{adjustbox}{max width=\textwidth}
  \begin{threeparttable}
    \caption{Bacterial Out-of-Distribution Performance}
    \label{tab:bacteria}
    \setlength{\tabcolsep}{4.5pt}
    \begin{tabular}{@{}l *{6}{c}@{}}
      \toprule
      Model & \makecell{Codon match\\(\%) $\uparrow$} & \makecell{$|\Delta \mathrm{GC}|$\\(pp) $\downarrow$} & \makecell{$|\Delta \mathrm{CFD}|$\\(pp) $\downarrow$} & \makecell{$|\Delta \mathrm{CSI}|$\\$\downarrow$} & \makecell{Normalized\\DTW $\downarrow$} & \makecell{$|\Delta \mathrm{MFE}|$\\(kcal/mol) $\downarrow$} \\
      \midrule
      \name (Ours) & \textbf{48.13} & \textbf{4.260} & \textbf{5.550} & \textbf{0.0887} & \textbf{8.219} & \textbf{28.97} \\
      CodonTransformer & 43.09 & 8.980 & 8.252 & 0.1308 & 19.875 & 55.44 \\
      SynCodonLM & 44.60 & 13.382 & 11.791 & 0.1885 & 29.405 & 77.41 \\
      \name (no MSA) & 38.89 & 6.247 & 14.060 & 0.2111 & 16.801 & 46.92 \\
      MSA signal only (no learning) & 47.39 & 5.086 & 5.766 & 0.0946 & 8.753 & 36.38 \\
      \bottomrule
    \end{tabular}
    \begin{tablenotes}[flushleft]
      \footnotesize
      \item Results are for 1,000 held-out bacterial targets across one run
    \end{tablenotes}
  \end{threeparttable}
  \end{adjustbox}
\end{table*}


\section{Conclusions and Future Work}

We develop an evolution-aware model translating protein sequences to nucleic acid sequences across species. Due to its grounding in biological knowledge, \name generates nucleic acid sequences that are substantially closer to the native sequence than traditional language models, with 17.95pp more matching codons. \name is a biologically-aware model which can translate across multiple biological modalities: amino acids, nucleic acids, and species. We demonstrate that providing evolutionary context provides a strong foundation for our model's predictions, and the flexibility of the model captures nuanced dependencies between the sequence, the evolutionary context, and the species context. 

Future work includes adapting this model for multi-objective gene-transfer design. Gene design objectives are complex and application-dependent. Our model generates a native-like nucleic acid sequence for a given protein and species; future work can explore using these sequences as a starting point. A natural extension would be to leverage fine-tuning or reinforcement learning to optimize the generated sequences for specific functional properties, including solubility and expression, with wet-lab validation. 

More broadly, our work emphasizes that explicitly encoding biological knowledge, rather than relying on scale to recover it, is a productive design principle for biological language models. Future work could incorporate biological signals beyond evolution and include additional modalities. 

\subsection*{AI use statement}

In this work, we used generative AI tools to implement methods, create software code, assist in data formatting, to edit the research paper to improve readability, and to assist in drafting the abstract and appendix. In addition to a manual literature review, we used generative AI tools to identify relevant literature and we verified the accuracy of all citations. We take responsibility for the final content of this work, including text, claims or artifacts produced with the aid of generative AI.

\subsection*{Reproducibility statement}

We provide details of the methodology for reproducibility in our methods and our appendix. Our data is derived from publicly available datasets;we describe our procedure to obtain the data in the appendix. Additionally, we will release a publicly available codebase upon publication. 

\subsection{Acknowledgements}
This work was funded by the DOE Center for Advanced Bioenergy and Bioproducts Innovation
(U.S. Department of Energy, Office of Science, Biological and Environmental Research Program
under Award Number DE-SC0018420). Any opinions, findings, and conclusions or recommendations
expressed in this publication are those of the author(s) and do not necessarily reflect the views of the U.S. Department of Energy.
This research used  the DeltaAI advanced computing and data resource, which is supported by the National Science Foundation (award OAC 2320345) and the State of Illinois. Delta and DeltaAI are joint efforts of the University of Illinois Urbana-Champaign and its National Center for Supercomputing Applications.

\bibliography{iclr2027_conference}
\bibliographystyle{iclr2027_conference}
\clearpage
\appendix
\section{Appendix}\label{sec:appendix}

\subsection{Dataset Construction and Quality Control}
\label{app:data}

\subsubsection{Ensembl gene trees and sequence filtering}

We constructed the primary dataset from the default aligned protein gene-tree
collections in Ensembl Compara release 116 and Ensembl Genomes release 63
\citep{yates2026ensembl}. We
considered protein members with canonical gene membership, a stable tree
identifier, and an associated alignment. Protein sequences were obtained from
the Compara sequence tables, coding sequences (CDSs) from the corresponding
member-sequence tables, and alignment coordinates from each member's CIGAR
string.

We retained only CDSs containing the canonical bases \texttt{A}, \texttt{C},
\texttt{G}, and \texttt{T}. Each CDS was required to translate under the
standard nuclear genetic code to its recorded peptide, to contain no internal
stop codon, and to be consistent with its alignment CIGAR. A terminal stop was
retained as a separate prediction position when present. Species identifiers
were normalized to their nearest NCBI \citep{schoch2020ncbi} species-rank ancestor, resulting in 882
normalized species. When a gene tree contained more than one valid sequence
from the same normalized species, we retained the lexicographically smallest
division/genome/protein identifier. We reapplied the requirement of at least
25 valid sequences from at least 25 distinct normalized species after all
quality-control and representative-selection steps. 

The protein alignment supplied by Ensembl defines the correspondence between
rows. We back-translated each aligned amino-acid residue to its native CDS
codon; the nucleotide alignment was therefore not constructed independently.
For a particular target, columns in which the target was a gap were omitted,
so the resulting feature sequence has one position per target codon.

\subsubsection{Data Splits}

Table~\ref{tab:appendix-split-sizes} gives the realized split sizes. 

\begin{table*}[htbp]
\centering
\small
\caption{Realized sequence and gene-tree counts for the three reported repeated splits.}
\label{tab:appendix-split-sizes}
\begin{tabular}{@{}lrrrrrrrr@{}}
\toprule
& \multicolumn{2}{c}{Training} & \multicolumn{2}{c}{Standard set validation}
& \multicolumn{2}{c}{Standard test} & \multicolumn{2}{c}{Within-tree test} \\
\cmidrule(lr){2-3}\cmidrule(lr){4-5}\cmidrule(lr){6-7}\cmidrule(l){8-9}
Run & Sequences & Trees & Sequences & Trees & Sequences & Trees & Sequences & Trees \\
\midrule
1 & 5,070,637 & 48,510 & 296,679 & 6,716 & 296,703 & 6,708 & 269,705 & 48,510 \\
2 & 5,071,532 & 48,507 & 296,685 & 6,710 & 296,685 & 6,717 & 268,822 & 48,507 \\
3 & 5,071,019 & 48,513 & 296,691 & 6,709 & 296,666 & 6,712 & 269,348 & 48,513 \\
\bottomrule
\end{tabular}
\end{table*}

\subsubsection{Synonymous-codon conservation analysis}
\label{app:conservation}

For the conservation analysis, we excluded gaps, stop codons, methionine, and
tryptophan. Within each alignment column, residues were partitioned by amino
acid, and an amino-acid group contributed only when at least two comparable
residues were present. Every unordered pair in an eligible group contributed
to the denominator, and a pair was counted as conserved when both members used
the same synonymous codon. If multiple amino-acid groups were eligible in the
same column, their matching-pair and total-pair counts were pooled before
calculating the column score. Reported conservation is the equal-weight mean
of these per-column scores.

The null expectation preserves the gene tree, alignment column, species,
amino acids, and gaps. For species $s$ and $t$ in an amino-acid group,
the expected probability of a synonymous match is calculated analytically as
\[
P_{0}(\text{match}\mid a,s,t)
=
\sum_{c\in\mathcal C(a)}\pi_{s,c}\pi_{t,c},
\]
where the species-specific distributions for this descriptive analysis were
calculated over the complete dataset. Thus, the null does not assume a
uniform codon distribution. Confidence intervals were obtained
from 2,000 gene-tree bootstrap replicates stratified by Ensembl division.

\subsubsection{Bacterial Out-of-Distribution Dataset}
\label{app:bacteria}

We selected bacterial genomes from GTDB release R232
\citep{parks2026gtdb}. Eligible assemblies were GTDB species representatives
with estimated completeness of at least 90\%, contamination of at most 5\%,
and NCBI \citep{sayers2025database} translation table 11. We ranked phyla by the number of eligible
genera and selected the ten largest phyla, breaking ties by phylum name. Within
each phylum, genomes were first ranked within genus by isolate status,
completeness, contamination, assembly level, and accession. A deterministic
SHA-256 ordering with seed 232001 then selected 50 distinct genera per phylum,
giving 500 genomes.

Only the selected assemblies were downloaded. Protein-coding genes were
called with Prodigal 2.6.3 \citep{hyatt2010prodigal} in single-genome mode with
translation table 11. Partial genes and CDSs that failed canonical translation
to their reported proteins were removed. OrthoFinder
\citep{emms2019orthofinder} was run separately within each phylum using
\texttt{diamond\_ultra\_sens}. We retained orthogroups with exactly one protein
per occupied genome and at least 25 occupied species. Each candidate group was
aligned at the protein level with MAFFT \citep{katoh2013mafft}, after which the
corresponding CDSs were back-translated through the amino-acid alignment.

Candidate targets were divided into length strata of at most 250, 251--500,
and more than 500 amino acids. We selected 100 primary orthogroups per phylum
and retained up to 20 reserves per phylum. Targets were chosen using a seeded
ordering that balanced assignments across genomes, with at most five targets
per genome. Each selected target came from a different orthogroup. Targets
were limited to 2,047 sense amino acids so that the appended terminal-stop
position produced at most 2,048 codons. The resulting benchmark contains
exactly 1,000 targets, with 100 from each of ten phyla. The target is excluded
from its own context, leaving 24--49 homolog rows. At a target position, a
homolog contributes only when its aligned residue encodes the same amino acid;
gaps and different amino acids do not contribute.

To exclude overlap with model training, each candidate target was searched
against the 5,070,637 Run~1 training proteins using MMseqs2 \citep{steinegger2017mmseqs2} with
\texttt{--min-seq-id 0.8 -c 0.8 --cov-mode 0 -s 7.5}. A target was rejected if
any training hit had at least 80\% identity, 80\% query coverage, and 80\%
target coverage; a reserve from the same phylum and length stratum was then
used.

\subsection{Architecture Details}\label{app:model}

The amino-acid vocabulary contains the 20 standard amino acids, a terminal-
stop state, and padding. The codon-input vocabulary contains the 64 native
codons, one universal mask token, and padding. For each biological position,
we sum its amino-acid embedding, codon-or-mask embedding, projected
34-dimensional context, and fixed sinusoidal positional encoding. The context
projection is a linear map from 34 to 768 dimensions followed by layer
normalization, GELU, and dropout with probability 0.1. A learned species
embedding is prepended to the biological sequence. Each run's vocabulary is
constructed from training species, with identifier 0 reserved for an unknown
species; the reported runs contain 882 training-species tokens plus the
unknown token.

The encoder contains 13 pre-layer-normalized Transformer blocks, hidden width
768, 12 attention heads of width 64, and a 2,944-dimensional GELU \citep{hendrycks2016gaussian} feed-forward
sub-layer. Dropout is 0.1. The implementation uses FlashAttention 2 \citep{dao2024flashattention}(version
2.8.3 in the recorded runs) and a final layer normalization. The six-dimensional
output head represents amino-acid-specific synonymous-codon slots. Before both
loss calculation and decoding, invalid slots are assigned a large negative
logit. The model contains 90,365,062 trainable parameters.

\subsection{Training}
\label{app:training}

During the first four epochs, each eligible position is independently selected
with probability 0.15 and is replaced with the universal codon-mask
token, with training details in Table \ref{tab:appendix-ourmodel-training}. Sequences longer than
2,048 biological positions are truncated to their first 2,048 positions during
training.

At inference, all target codons are replaced with the universal mask token in
a single non-autoregressive pass. The synonymous codon with the largest valid
logit is selected deterministically at every position. Sequences with length greater than 2,048 are
partitioned into consecutive, non-overlapping chunks of at most 2,048
positions. The species token is repeated, positional indices restart, and
each chunk is decoded independently before concatenation. 

\begin{table*}[t]
\centering
\small
\caption{Training configuration for \name.}
\label{tab:appendix-ourmodel-training}
\begin{tabular}{@{}lll@{}}
\toprule
Setting & Epochs 1--4 & Sequence-property fine-tuning \\
\midrule
Target masking & 15\% & 100\% \\
Loss & Masked natural CE & Natural CE $+\mathcal L_{\mathrm{property}}$ \\
Optimizer & AdamW & AdamW \\
Peak learning rate & $2\times10^{-4}$ & $2\times10^{-6}$ \\
Schedule & Constant & 250-step warmup, cosine decay \\
Minimum learning rate & $2\times10^{-4}$ & $2\times10^{-7}$ \\
Weight decay & 0.01 & 0.01 \\
Gradient clipping & 1.0 & 1.0 \\
Precision & FP16 mixed & FP16 mixed \\
Maximum biological positions & 2,048 & 2,048 \\
Padded tokens per optimizer update & approximately 98,304 & approximately 98,304 \\
Checkpoint selection & Completed epoch 4 & Joint fully masked validation objective \\
\bottomrule
\end{tabular}
\end{table*}

\subsection{Results on sets matching SynCodonLM}\label{app:syncodonset}
Since we could only run 50,000 sequences in the standard test set and 10,000 sequences in the within-tree test set with length no greater than 2,048 on SynCodonLM \citep{heuschkel2026advancing} due to long inference time, we additionally provide results of all models on only these 50,000 and 10,000 sequences, respectively, in Tables \ref{tab:ourmodel-ablation-comparison-whole-tree-50k} and \ref{tab:ourmodel-ablation-comparison-within-tree-10k}. 

\begin{table*}[t]
  \centering
  \small
  \begin{adjustbox}{max width=\textwidth}
  \begin{threeparttable}
    \caption{Comparison on a Shared 50,000-Target Cohort (standard dataset)}
    \label{tab:ourmodel-ablation-comparison-whole-tree-50k}
    \setlength{\tabcolsep}{4.5pt}
    \begin{tabular}{@{}l *{6}{c}@{}}
      \toprule
      Model & \makecell{Codon match\\(\%) $\uparrow$} & \makecell{$|\Delta \mathrm{GC}|$\\(pp) $\downarrow$} & \makecell{$|\Delta \mathrm{CFD}|$\\(pp) $\downarrow$} & \makecell{$|\Delta \mathrm{CSI}|$\\$\downarrow$} & \makecell{Normalized\\DTW $\downarrow$} & \makecell{$|\Delta \mathrm{MFE}|$\\(kcal/mol) $\downarrow$} \\
      \midrule
      \name (Ours) & \textbf{68.01 $\pm$ 0.06} & \textbf{1.422 $\pm$ 0.007} & \textbf{0.961 $\pm$ 0.008} & \textbf{0.0210 $\pm$ 0.0001} & \textbf{4.533 $\pm$ 0.010} & \textbf{17.95 $\pm$ 0.12} \\
      CodonTransformer & 50.03 $\pm$ 0.10 & 6.852 $\pm$ 0.478 & 3.210 $\pm$ 0.065 & 0.1189 $\pm$ 0.0000 & 14.240 $\pm$ 0.429 & 73.39 $\pm$ 4.78 \\
      SynCodonLM & 43.21 $\pm$ 1.30 & 11.055 $\pm$ 1.377 & 3.812 $\pm$ 0.353 & 0.1237 $\pm$ 0.0152 & 19.089 $\pm$ 2.492 & 104.00 $\pm$ 19.95 \\
      \name (no MSA ablation) & 50.37 $\pm$ 0.06 & 2.848 $\pm$ 0.044 & 1.504 $\pm$ 0.007 & 0.0408 $\pm$ 0.0006 & 6.400 $\pm$ 0.031 & 31.36 $\pm$ 0.89 \\
      MSA signal only (no learning) & 63.07 $\pm$ 0.05 & 2.696 $\pm$ 0.010 & 1.426 $\pm$ 0.017 & 0.0352 $\pm$ 0.0003 & 5.360 $\pm$ 0.026 & 24.91 $\pm$ 0.12 \\
      \bottomrule
    \end{tabular}
  \end{threeparttable}
  \end{adjustbox}
\end{table*}

\begin{table*}[t]
  \centering
  \small
  \begin{adjustbox}{max width=\textwidth}
  \begin{threeparttable}
    \caption{Within-tree Comparison on a Shared 10,000-Target Cohort}
    \label{tab:ourmodel-ablation-comparison-within-tree-10k}
    \setlength{\tabcolsep}{4.5pt}
    \begin{tabular}{@{}l *{6}{c}@{}}
      \toprule
      Model & \makecell{Codon match\\(\%) $\uparrow$} & \makecell{$|\Delta \mathrm{GC}|$\\(pp) $\downarrow$} & \makecell{$|\Delta \mathrm{CFD}|$\\(pp) $\downarrow$} & \makecell{$|\Delta \mathrm{CSI}|$\\$\downarrow$} & \makecell{Normalized\\DTW $\downarrow$} & \makecell{$|\Delta \mathrm{MFE}|$\\(kcal/mol) $\downarrow$} \\
      \midrule
      \name (Ours) & \textbf{65.67 $\pm$ 0.02} & \textbf{1.453 $\pm$ 0.014} & \textbf{0.951 $\pm$ 0.004} & \textbf{0.0216 $\pm$ 0.0002} & \textbf{4.608 $\pm$ 0.020} & \textbf{21.30 $\pm$ 0.09} \\
      CodonTransformer & 50.04 $\pm$ 0.45 & 6.678 $\pm$ 0.598 & 3.146 $\pm$ 0.087 & 0.1100 $\pm$ 0.0006 & 12.306 $\pm$ 0.353 & 86.81 $\pm$ 6.17 \\
      SynCodonLM & 42.78 $\pm$ 1.04 & 11.153 $\pm$ 1.431 & 4.073 $\pm$ 0.383 & 0.1247 $\pm$ 0.0131 & 18.167 $\pm$ 2.335 & 122.97 $\pm$ 21.78 \\
      \name (no MSA ablation) & 50.58 $\pm$ 0.19 & 2.610 $\pm$ 0.025 & 1.388 $\pm$ 0.009 & 0.0380 $\pm$ 0.0003 & 5.952 $\pm$ 0.029 & 35.07 $\pm$ 1.46 \\
     MSA signal only (no learning) & 58.41 $\pm$ 0.06 & 3.742 $\pm$ 0.011 & 2.032 $\pm$ 0.070 & 0.0482 $\pm$ 0.0008 & 6.334 $\pm$ 0.019 & 37.18 $\pm$ 0.40 \\
      \bottomrule
    \end{tabular}
  \end{threeparttable}
  \end{adjustbox}
\end{table*}

\subsection{Comparison Methods}
\label{app:comparators}

\subsubsection{CodonTransformer}

We followed the official CodonTransformer implementation
\citep{fallahpour2025codontransformer} at Git commit
\texttt{4a447b01dab860feb81b647ff1ff88ad598517f4}. We adapted its data-loading
pipeline to train on our three dataset splits, constructed the
species vocabulary using only the training species from each split, and used
deterministic validation loss for checkpoint selection. To avoid
discarding positions from sequences longer than the model context window, we
divided them into consecutive chunks of at most 2,046 codons and concatenated
their predictions. 

\subsubsection{SynCodonLM}

We followed the official SynCodonLM implementation
\citep{heuschkel2026advancing} at Git commit
\texttt{9a773d9e44d1ea6e94e872cbabf41161edb54a28}. We adapted its data-loading
pipeline to train independently on our three dataset splits, divided sequences
longer than its context window into consecutive non-overlapping windows, and
used deterministic validation loss for checkpoint selection. For a
matched training budget, we trained for the same five epochs used by our method.
To make training and inference practical on our hardware, we vectorized the
synonymous masked-token loss, and batched independent decoding windows.

\end{document}